\documentclass[
    ,final            
  ]
  {aipproc}

\layoutstyle{6x9}

\usepackage{lscape,chngpage}

\usepackage{amsmath,amssymb,rotating,wasysym,mathrsfs}

\usepackage{graphicx,tikz}

\def\Journal#1#2#3#4{{#1} {\bf #2}, #3 (#4)}

\def\CQG{\em Class. Quantum Grav.}
\def\PRD{\em Phys. Rev. D }
\def\GRG{\em Gen. Rel. Grav.}

\def\JMP{\em J. Math. Phys.}

\def\PRL{\em Phys. Rev. Lett.}

\def\be{\begin{equation}}
\def\ee{\end{equation}}
\def\bea{\begin{eqnarray}}
\def\eea{\end{eqnarray}}
\def\bean{\begin{eqnarray*}}
\def\eean{\end{eqnarray*}}

\def\espaitemps{({\cal V},g)}
\def\varietat{{\cal V}}

\def\unwarrow{\nwarrow\joinrel\joinrel\uparrow}
\def\unearrow{\uparrow\joinrel\joinrel\nearrow}

\def\dotsearrow{\hspace{-5mm}\dot{\hspace{5mm}\searrow}}

\def\dotdswarrow{\swarrow\joinrel\joinrel\dot\downarrow}
\def\dotdsearrow{\dot\downarrow\joinrel\joinrel\searrow}

\def\unarrow{\nwarrow\joinrel\joinrel\uparrow\joinrel\joinrel\nearrow}

\def\xiv{\vec \xi }
\def\etav{\vec \eta}
\def\lie{{\pounds}}

\def\AH{\mbox{AH}}

\def\scri{\mathscr{J}}
\def\B{\mathscr{B}}
\def\R{\mathscr{R}}

\def\r{\R_0}

\begin{document}

\title{On the boundary of the region containing trapped surfaces}


\author{Jos\'e M. M. Senovilla}{
  address={F\'{\i}sica Te\'orica, Universidad del Pa\'{\i}s Vasco, Apartado 644, 48080 Bilbao, Spain.}
}



\begin{abstract}
The boundary of the region in spacetime containing future-trapped closed surfaces is considered. In asymptotically flat spacetimes, this boundary does not need to be the event horizon nor a dynamical/trapping horizon. Some properties of this boundary and its localization are analyzed, and illustrated with examples. In particular, fully explicit future-trapped compact surfaces penetrating into flat portions of a Vaidya spacetime are presented.
\end{abstract}

\maketitle


\section{Introduction}
In this contribution I would like to address the following question:
what is the surface of an evolving black hole? Concentrating on the case of asymptotically flat black holes \cite{Haw,HE}, the standard candidate is the event horizon (EH). Unfortunately, EH suffers from a serious problem: it is teleological, depending on the whole future evolution of the spacetime. However, if a black hole is evolving or forming, we would like to know how to recognize it ---and we, of course, do not control nor know the entire future evolution of the spacetime. 

This led to the definition of some quasi-local objects, essentially the {\em future outer trapping horizons} \cite{Hay,Hay1} and the {\em dynamical horizons} \cite{AK1}, in order to characterize the boundary of asymptotically flat black holes, see \cite{AK1,AG,B,Kri} and references therein. 
Both of these quasi-local objects are spacelike {\em marginally trapped tubes} \cite{AG}: hypersurfaces foliated by marginally future-trapped closed surfaces. It turns out, however, that these quasi-local horizons are not unique in general \cite{AG}. Even more problematic, they do not separate regions with and without future-trapped surfaces, as follows from the result in \cite{BS}: closed future-trapped surfaces can penetrate flat regions of imploding Vaidya spacetimes. This will be summarized in the next section. Thus the quasi-local horizons are of limited use concerning the question under study.

On the other hand, Eardley \cite{E} conjectured that the EH is the boundary of the set of marginally {\em outer} future-trapped closed surfaces ---these are compact surfaces with vanishing outer expansion, see e.g. \cite{AMS,AMS1}. This conjecture holds true for the imploding Vaidya spacetimes \cite{BD}. An obvious question arises: what is the boundary $\B$ of the set of truly future-trapped closed surfaces? $\B$ is a boundary enclosing the region {\em where} dynamical and future outer trapping horizons can exist. It follows that $\B$ is a good candidate for the sought "surface of a black hole", for it is the genesis of quasi-local horizons and f-trapped closed surfaces.

Hence, $\B$ is the main object to be analyzed herein. In \cite{BD} it was proven that $\B\neq$ EH in general. Moreover, from the mentioned result in \cite{BS} follows that $\B$ does penetrate flat portions of imploding black hole spacetimes, which immediately rules out the quasi-local horizons as candidates for $\B$. Actually, $\B$ cannot contain any marginally future-trapped closed surface. This, together with other results for $\B$ found in \cite{BS1}, will be presented here. Even though the techniques can be used in general situations, I will concentrate on the case of spherical symmetry. For this case, very definite limits on the location of $\B$ will be given. For the exact characterization ---and precise location---of $\B$ in spherical symmetry readers are referred to \cite{BS1}.

\subsection{Preliminaries and notation: the trapped surface fauna}
Let $\espaitemps$ be a 4-dimensional causally orientable 
spacetime with metric $g_{\mu\nu}$ of signature $-,+,+,+$. 
Let $S$ be a connected 2-dimensional
surface with local intrinsic coordinates $\{\lambda^A\}$ imbedded in
$\varietat$ by the smooth parametric equations
$x^{\alpha} =\Phi ^{\alpha}(\lambda^A) $
where $\{x^{\alpha}\}$ are local coordinates for $\varietat$.
The tangent vectors $\vec{e}_A$ of $S$ are locally given by
\bean
\vec{e}_A \equiv e^{\mu}_A \left.\frac{\partial}{\partial x^{\mu}}
\right\vert _S \equiv
\frac{\partial \Phi^{\mu}}{\partial \lambda^A}
\left.\frac{\partial}{\partial x^{\mu}}\right\vert_S
\eean
so that the first fundamental form of $S$ in $\varietat$ is: $
\gamma _{AB}\equiv \left.g_{\mu \nu}\right\vert_S\frac{\partial \Phi^{\mu}}
{\partial \lambda^A}\frac{\partial \Phi^{\nu}}{\partial \lambda^B}$.
We assume that $S$ is spacelike ergo $\gamma _{AB}$ is positive definite. The two linearly independent one-forms
$k_{\mu}^{\pm}$ normal to $S$ can be chosen to be null and future directed
everywhere on $S$, so they satisfy
$$
k_{\mu}^{\pm}e^{\mu}_A=0,\,  k^{+}_{\mu} k^{+ \mu}=0 ,\,
 k^{-}_{\mu} k^{- \mu}=0 ,\hspace{3mm}  k_{\mu}^+ k^{-\mu}=-1 . 
$$
The last equality is a condition of normalization despite which there remains the freedom
\begin{equation}
k^+_{\mu} \longrightarrow k'^+_{\mu}=\sigma^2 k^+_{\mu}, \hspace{1cm}
k^-_{\mu} \longrightarrow k'^-_{\mu}=\sigma^{-2} k^-_{\mu} \label{free}
\end{equation}
where $\sigma^2$ is a positive function defined on $S$.
The orthogonal splitting into directions tangential or normal 
to $S$ leads to the standard formula \cite{Kr,O}:
$$
\nabla_{\vec{e}_{A}}\vec{e}_{B}=\overline{\Gamma}^{C}_{AB}\vec{e}_{C}-\vec{K}_{AB}
$$
where $\overline{\Gamma}^{C}_{AB}$ are the symbols of the Levi-Civita 
connection $\overline\nabla$ of 
$\gamma$
and $\vec{K}_{AB}$ is the shape tensor of $S$ in $\espaitemps$. Observe that $\vec{K}_{AB}=\vec{K}_{BA}$ and it is
orthogonal to $S$, so that we can write
$$
\vec{K}_{AB}=-K^-_{AB}\, \vec k^+ -K^+_{AB}\, \vec k^- \, .
$$
$K^{\pm}_{AB}$ are the two null (future) second fundamental forms of $S$ in $\espaitemps$, defined by
$$
K^{\pm}_{AB} \equiv -k^{\pm}_{\mu}e^{\nu}_A\nabla_{\nu}e^{\mu}_B = e^{\mu}_Be^{\nu}_A\nabla_{\nu}k^{\pm}_{\mu} .
$$
The shape tensor enters in the fundamental relation
\be
e^{\mu}_{A}e^{\nu}_{B}\nabla_{\mu}v_{\nu}|_{S}=\overline\nabla_{A} 
\overline{v}_{B}+v_{\mu}|_{S} K^{\mu}_{AB}
\label{nablas2}
\ee
where, for all $v_{\mu}$ we denote by $\overline{v}_{B}\equiv v_{\mu}|_{S}\,\, e^{\mu}_{B}$
its projection to $S$.

The mean curvature vector of $S$ in $\espaitemps$ \cite{O,Kr} is defined as
$$
\vec{H}\equiv \gamma^{AB}\vec{K}_{AB}
$$
where $\gamma^{AB}$ is the contravariant metric on $S$. $\vec H$ is orthogonal to $S$, invariant under transformations (\ref{free}) and 
$$
\vec{H}= -\theta^-\vec{k}^+ - \theta^+\vec{k}^- , \hspace{2cm}
\theta^{\pm} \equiv \gamma^{AB}K^{\pm}_{AB} .
$$
$\theta^\pm$ are called the (future) null expansions.

The class of {\em generically} future trapped (f-trapped from now on) surfaces are characterized by having $\vec H$ pointing to the future everywhere on $S$, and similarly for past 
trapped. These conditions can be equivalently expressed in terms of the signs of the expansions: $\theta^\pm\leq 0$.
A convenient way of visualizing the possible cases is achieved by using an arrow notation for $\vec H$ and the convention that upwards means ``future" and 45$^o$ means ``null".
The full list of possibilities for generically f-trapped surfaces is collected in the next table where the symbol is defined by the causal orientation(s) of $\vec H$: see \cite{S4} for further details and a refined classification.\footnote{This is to be compared with \cite{Wald,AG,HE}, as sometimes different names are given to the same objects, and vice versa, different objects are called with the same name.}
\begin{center}
\begin{tabular}{c|c|l}
$\vec{H}$-orientation & Expansions & Type of surface \\
\hline
$\nearrow$ & $\theta^+=0, \theta^-<0$ & marginally f-trapped \\
$\nwarrow$ & $\theta^+<0, \theta^-=0$ & marginally f-trapped \\
$\uparrow$ & $\theta^+<0, \theta^-<0$ & f-trapped\\
\begin{sideways}{$\dotsearrow$}\end{sideways} & $\theta^+=0, \theta^-\leq 0$ & partly marginally f-trapped \\
\begin{sideways}\begin{sideways}{$\dotsearrow$}\end{sideways}\end{sideways} & $\theta^+\leq 0, \theta^-=0$ & partly marginally f-trapped \\
\begin{sideways}\begin{sideways}$\dot\downarrow$\end{sideways}\end{sideways} & $\theta^+\leq0, \theta^-\leq0$ ($\theta^+=0\Leftrightarrow \theta^-=0$) & partly f-trapped\\
$\unearrow$ & $\theta^+\leq 0, \theta^-<0$ & almost f-trapped\\
$\unwarrow$ & $\theta^+<0, \theta^-\leq 0$ & almost f-trapped\\
\begin{sideways}\begin{sideways}$\swarrow\dotsearrow$\end{sideways}\end{sideways} & 
$\theta^+\leq 0, \theta^-\leq 0$, $\theta^+\theta^-=0$ &
null f-trapped\\
\begin{sideways}\begin{sideways}$\dotdswarrow$\end{sideways}\end{sideways} & 
$\theta^+\leq 0, \theta^-\leq 0$ ($\theta^-=0\Rightarrow \theta^+=0$) & feebly f-trapped \\ \begin{sideways}\begin{sideways}$\dotdsearrow$\end{sideways}\end{sideways} &
$\theta^+\leq 0, \theta^-\leq 0$ ($\theta^+=0\Rightarrow \theta^-=0$) & feebly f-trapped \\
$\unarrow$ & $\theta^+\leq 0, \theta^-\leq 0$ ($\theta^++\theta^-<0$) & weakly f-trapped \\
\begin{sideways}\begin{sideways}$\swarrow\dotsearrow$
\end{sideways}\end{sideways}\hspace{-7.1mm} \raisebox{2mm}{$\uparrow$}\hspace{2mm} & $\theta^+\leq 0, \theta^-\leq 0$ & nearly f-trapped \\
\hline
\end{tabular}
\end{center}
Here the $\cdot$ indicates the cases with $\vec H =\vec 0$. The important case of minimal surfaces has $\theta^+=\theta^-=0$ (that is $\vec H =\vec 0$) everywhere on $S$ and can be considered as a limit case. 

\section{Closed trapped surfaces penetrate flat regions of imploding Vaidya spacetimes}
Consider the Vaidya spacetime with incoming radiation, whose line-element is \cite{V,Exact}
\be
ds^2=-\left(1-\frac{2m(v)}{r}\right)dv^2+2dvdr+r^2d\Omega^2 \label{ds2}
\ee 
where $d\Omega^2$ is the standard metric on the unit round spheres, $v$ is radial null advanced time and $m(v)\geq 0$ is the mass function. The Einstein tensor of (\ref{ds2}) takes the form
$$
G_{\mu\nu}=\frac{2}{r^2}\frac{dm}{dv}\ell_\mu\ell_\nu
$$
where the null vector field 
$$\vec\ell =-\partial_r , \hspace{1cm} \ell_\mu dx^\mu=-dv\hspace{6mm} (\ell^\mu\ell_\mu=0)
$$
is future pointing. Hence, if Einstein's field equations are assumed, the energy conditions \cite{HE} imply
\be
\frac{dm}{dv}\geq 0 .\label{mdot}
\ee

The preferred 2-spheres (defined by constant values of $v$ and $r$) have the following null expansions
$$
\theta^+ =\frac{1}{2r} \left(1-\frac{2m(v)}{r}\right), \hspace{3mm} \theta^-=-\frac{1}{r}
$$
where $\vec k^-=\vec\ell$ and $\vec k^+ =\partial_v+(1/2-m(v)/r) \partial_r$. Hence, they are (marginally) f--trapped if and only if $r<2m(v)$ ($r=2m(v)$). The hypersurface defined by
$$
\mbox{AH:} \hspace{2mm} r-2m(v)=0
$$
is foliated by marginally f-trapped 2-spheres. It is called the spherically symmetric  ``apparent 3-horizon". It can be checked that AH is a spacelike hypersurface whenever $dm/dv>0$, and it is null where $dm/dv=0$. Therefore, AH is a dynamical horizon \cite{AK1} as well as a future outer trapping horizon \cite{Hay,AG} on the region where it is spacelike ---and an isolated horizon \cite{AK1} where $m(v)=$const. 

The analysis will be restricted to cases with a continuous piecewise differentiable $m(v)$ such that
\be
m(v)=0 \hspace{3mm} \forall v<0; \hspace{1cm} m(v)\leq M <\infty \hspace{3mm} \forall v>0
\label{mass}
\ee
together with (\ref{mdot}), where $M$ is a constant (the final total mass). These Vaidya spacetimes tend asymptotically to the Schwarzschild solution with total mass $M$. Observe, on the other hand, that the spacetime is {\em flat} for the entire portion with $v<0$.

\begin{figure}[!ht]
\includegraphics[width=8.6cm]{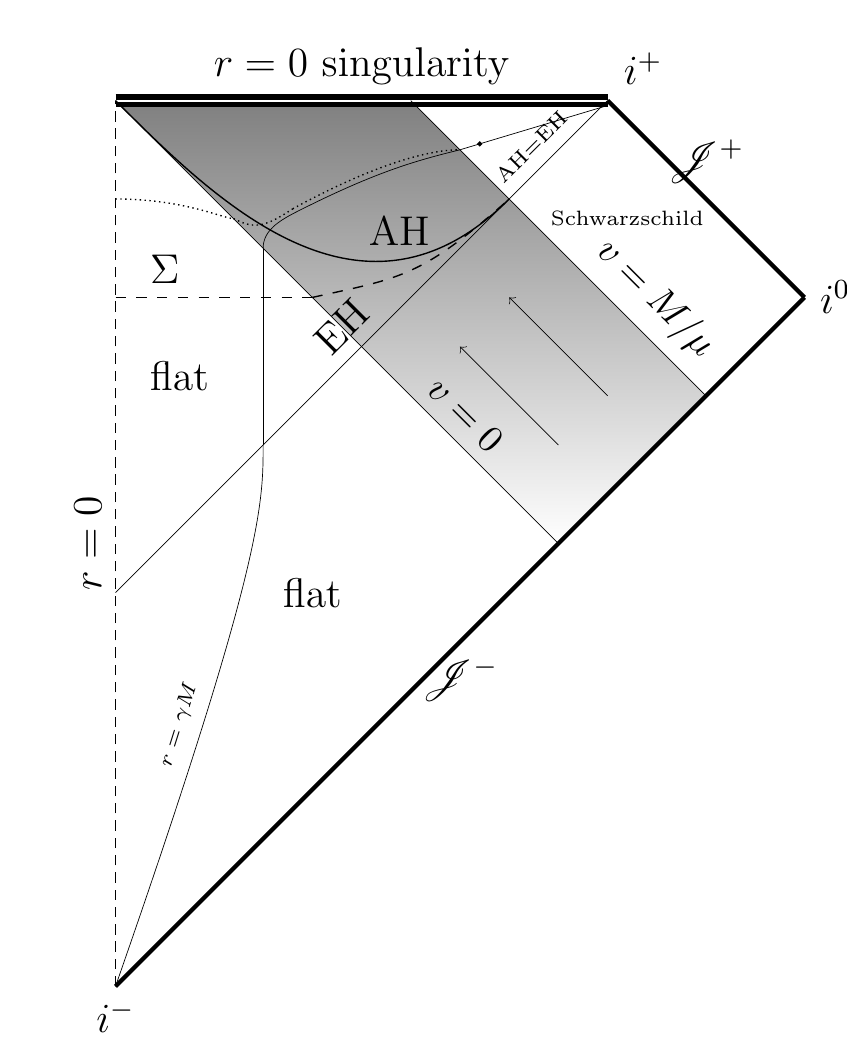}
\caption{Conformal diagram of spacetime (\ref{ds2}) with (\ref{mass2}). The discontinuous line marked as $r=0$ is the origin of coordinates. A dressed curvature singularity is present at $r=0,\, v>0$. This is a spacelike future singularity. The spacetime is initially flat until null radiation flows in spherically from past infinity starting at $v=0$. The shaded region is a pure self-similar Vaidya spacetime. The spacetime becomes Schwarzschild with mass $M$ for all $v>M/\mu$. The apparent 3-horizon AH is spacelike in the shaded Vaidya region and then merges with EH at the 2-sphere $v=M/\mu$, $r=2M$. Notice that EH starts developing in the flat region. The portion above EH but below $v=0$ is a causal diamond in flat spacetime ---the intersection of two light cones--- and therefore it can de drawn without conformal distortion. A f-trapped closed surface entering the flat region is explicitly constructed in the main text. In the diagram it is represented by the dotted line, with a final portion within the $r=\gamma M$ hypersurface, with $\gamma <0.68514$. Nevertheless, f-trapped closed surfaces cannot extend below the hypersurface $\Sigma$ represented by the dashed line, and they all must have a non-empty portion above the AH, as it is explained in the main text.}
\label{fig:fig1}
\end{figure}

The event horizon EH is a spherically symmetric null hypersurface due to its definition as $\partial J^-(\scri^+)$ where $\scri^+$ denotes future null infinity \cite{Haw,HE,Wald}. One has $r|_{EH}\geq 2m(v)$, the equality holding only if $m(v)=M$ for all $v>v_1$. It is important to realize that EH penetrates the flat portion of the spacetime --- a manifestation of its teleological character. The actual position of EH depends on the form of the mass function $m(v)$. To be specific, we are going to choose the following simple case \cite{HWE}
\be
m(v)=\left\{\begin{array}{cl}
0, & v\leq 0 \\
\mu v, & 0\leq v\leq M/\mu \\
M, & v\geq M/\mu
\end{array}
\right.\label{mass2}
\ee
where $\mu$ is a positive constant. This spacetime is self-similar in the non-empty region $0< v< M/\mu$, and describes the collapse of a finite shell of incoherent radiation entering flat spacetime in a spherically symmetric manner from the past, leading to a Schwarzschild black hole of mass $M$. To avoid the formation of naked singularities the restriction $\mu >1/16$ must be imposed \cite{HWE,Pap,K}. The Penrose diagram of this particular Vaidya spacetime is depicted in figure \ref{fig:fig1}.

Closed f-trapped surfaces cannot extend all the way up to the portion of EH in the flat region, as was proved in \cite{BD}. However, numerical investigations \cite{SK} were incapable of finding closed f-trapped surfaces to the past of the apparent 3-horizon AH. The resolution of whether or not f-trapped closed surfaces can penetrate into the flat region ---and then also cross the AH--- was solved only recently in \cite{BS}, where fully explicit examples are constructed. These surfaces are composed of the following parts:
\begin{itemize}
\item Flat region: a topological disk given by the hyperboloid
$$
\theta =\pi/2 ; \hspace{2cm} v=t_0 +r -\sqrt{r^2+k^2} 
$$
with constants $t_0,k$.
\item Vaidya self-similar region: a topological cylinder defined by $\theta =\pi/2$ and
$$
\sqrt{v^2-bvr+ar^2}=C\exp \left\{\frac{b}{2\sqrt{a-b^2/4}}\arctan\left(\frac{2v-br}{2r\sqrt{a-b^2/4}} \right)\right\}
$$
where $a,b$ and $C$ are constants subject to $a>b^2/4$. These have $dr/dv=0$ at $v=br$.
\item Schwarzschild region: another disk composed of two parts
\begin{itemize}
\item a cylinder with $\theta =\pi/2 ; \hspace{1mm} r=\gamma M$ where $\gamma$ is a positive constant.

\item another final ``capping" disk defined by
$$
\left(\theta -\frac{\pi}{2} +\delta \right)^2 +\left(\frac{v}{\gamma M}-c_1 \right)^2=\delta^2
$$
with  constants $c_1$ and $\delta$.
\end{itemize}
\end{itemize}
The total surfaces are topologically $S^2$, and they are future-trapped if 
$t_0<k$, $k>0$, $0<a<b$, $1>\gamma=(1/b\mu)$, $a\geq 1/\mu$, $0<\delta\leq \pi/2$ and $$\sqrt{\frac{2}{\gamma-1}}\left(\frac{1}{\gamma}-1\right) >\frac{1}{\delta}.$$
These conditions imply in turn a restriction on the growth of the mass function \cite{BS}:
$$
\mu=\frac{1}{\gamma b}>\frac{1}{\gamma}\frac{b}{4a}>\frac{1}{4\gamma}, \hspace{1cm} \gamma < 0.68514 .
$$
It should be remarked that this construction of an explicit f-trapped closed surface is not optimal in the given Vaidya spacetime, so that there may well be examples for smaller values of $\mu$. However, from the restrictions that we will impose later by means of the hypersurface $\Sigma$ ---see the next section---, f-trapped closed surfaces can never penetrate the flat region if $\mu \leq 1/8$.

The following conclusions are drawn from these results: (i) closed f-trapped surfaces can penetrate flat spacetime regions if the mass function rises fast enough; (ii) the dynamical horizon AH is not the sought for boundary $\B$ in general; and (iii) the teleological character of the EH translates into a {\em non-local} property of closed trapped surfaces, for they can have portions in a region of spacetime whose whole past is flat as long as some energy crosses them elsewhere to make their compactness feasible.

\section{Fundamental general results}
The main results leading to a better localization of the boundary $\B$ come from the interplay between (generalized) symmetries and generically trapped surfaces. They are fully general and based on ideas presented in \cite{MS,S1,S2,S3}.

We start with the identity $(\lie_{\xiv} g)_{\mu\nu}=\nabla_{\mu}\xi_{\nu}+\nabla_{\nu}\xi_{\mu}$ for arbitrary vector fields $\xiv$, where 
$\lie_{\xiv}$ denotes the Lie derivative with respect to $\xiv$. Projecting to $S$ and using (\ref{nablas2})
$$
(\lie_{\xiv} g|_{S})_{\mu\nu}\, e^{\mu}_Ae^{\nu}_B=\overline\nabla_{A}\overline\xi_{B}+
\overline\nabla_{B}\overline\xi_{A}+2 \xi_{\mu}|_{S} K^{\mu}_{AB}\, .
$$
Contracting now with $\gamma^{AB}$ we get the main formula to be exploited repeatedly in what follows
\be
\fbox{$\displaystyle{\frac{1}{2}
P^{\mu\nu}(\lie_{\xiv} g|_{S})_{\mu\nu} =
\overline\nabla_{C}\overline\xi^{C}+ \xi_\rho H^\rho}$}
\label{main}
\ee
where 
$$
P^{\mu\nu}\equiv \gamma^{AB} e^{\mu}_Ae^{\nu}_B
$$
is the orthogonal projector of $S$ ---it projects to the part tangent to $S$. 

This elementary formula (\ref{main}) is very useful. Observe, for instance, that if $S$ is compact without boundary
$$
\oint_{S}\xi_\rho H^\rho =\frac{1}{2} \oint_S P^{\mu\nu}(\lie_{\xiv} g|_{S})_{\mu\nu}
$$
so that the sign of $\xi_\rho H^\rho$ is related to the sign of the projection to $S$ of the deformation $(\lie_{\xiv} g)_{\mu\nu}$. Thus,
\begin{quotation}
{\em if $\xiv$ is future-pointing on a region $\R\subset \varietat$, then the closed $S$  cannot be contained in $\R$ and generically f-trapped} if $\oint_S P^{\mu\nu}(\lie_{\xiv} g|_{S})_{\mu\nu}\geq 0$. 
\end{quotation}
The only exception is the case where $S$ is (partly) marginally f-trapped, $\xiv|_S$ is null and proportional to $\vec H$ and $P^{\mu\nu}(\lie_{\xiv} g|_{S})_{\mu\nu} = 0$.

This general conclusion is applicable, for example, to {\em conformal Killing vectors} \cite{Exact} (including the homothetic and proper Killing vectors) and to {\em Kerr-Schild vector fields} \cite{CHS}. The former satisfy
\be
(\lie_{\vec\xi}g)_{\mu\nu}=2\psi g_{\mu\nu} \label{CKV}
\ee
for some function $\psi$, so that $P^{\mu\nu}(\lie_{\xiv} g|_{S})_{\mu\nu} =4\psi|_S$. Thus, the condition used in the reasoning above reduces to simply $\oint_S\psi \geq 0$. The Kerr-Schild vector fields are defined by
\be
(\lie_{\vec\xi}g)_{\mu\nu}=2h\ell_\mu \ell_\nu, \hspace{1cm} (\lie_{\vec\xi}\ell)_{\mu}=b\ell_\mu
\label{KSVF}
\ee
for some functions $h$ and $b$, where $\ell_\mu$ is a fixed {\em null} one-form field ($\ell_\mu\ell^\mu=0$). Therefore $P^{\mu\nu}(\lie_{\xiv} g|_{S})_{\mu\nu} =2h\, \bar\ell_A \bar\ell^A$ and the condition holds if $h|_S\geq 0$.

Consider now the case where $\xiv$ is hypersurface orthogonal
$$
\xi_{[\mu}\nabla_{\nu}\xi_{\rho]}=0 \hspace{2mm} \Longleftrightarrow \hspace{2mm} \xi_{\mu}=-F \partial_{\mu} \tau
$$
for some local functions $F>0$ and $\tau$. The hypersurfaces $\tau =$const. are called the level hypersurfaces and they are orthogonal to $\xiv$. 
\begin{quotation}
{\em Assume again that $\xiv$ is future-pointing on $\R\subset \varietat$, then any minimal or generically f-trapped surface $S$  cannot have a local minimum of $\tau$ at any point $q\in \R$ such that $P^{\mu\nu}(\lie_{\xiv} g)_{\mu\nu}|_q\geq 0$}. 
\end{quotation}
This follows because at any local minimum  $\bar\xi_A|_q=0$ from where one can derive using the main formula (\ref{main})
$$
\left.\bar F\gamma^{AB}\frac{\partial^2\bar\tau}{\partial\lambda^A\partial\lambda^B}\right|_q=\left.-\frac{1}{2}
P^{\mu\nu}(\lie_{\xiv} g)_{\mu\nu} \right|_q+
\left.\xi_\rho H^\rho\right|_q 
$$
so that $\partial^2\bar\tau/\partial\lambda^A\partial\lambda^B|_q$ cannot be positive (semi)-definite. A detailed complete proof is given in \cite{BS1}.

Some important remarks are in order here: first of all, observe that $S$ does not need to be compact, nor fully contained in $\R$. Letting aside the exceptional possibility of minimal surfaces contained in a $\tau =$constant hypersurface if they have $P^{\mu\nu}(\lie_{\xiv} g)_{\mu\nu}|_S=0$, this result implies that, under the stated conditions, one can always follow a connected path along $S\cap\R$ with decreasing $\tau$. Note, also, that the result applies in particular but not only to (i) {\em static} Killing vectors, (ii) hypersurface-orthogonal causal conformal Killing vectors (\ref{CKV}) with $\psi\geq 0$, and (iii) hypersurface-orthogonal causal Kerr-Schild vector fields (\ref{KSVF}) with $h\geq 0$.

Finally, consider the possibility of surfaces, compact or not, contained in one of the level hypersurfaces $\tau =$constant in $\R$. In that case, $\bar\xi_A=0$ all over $S\cap\R$ hence (\ref{main}) implies
$$
2\xi_\rho H^\rho =P^{\mu\nu}(\lie_{\xiv} g|_{S})_{\mu\nu}  .
$$
Thus, at any point $x\in S$ such that $P^{\mu\nu}(\lie_{\xiv} g)_{\mu\nu} |_x \geq 0$, $\vec H|_x$  cannot be timelike future-pointing, and it can be future-pointing null or zero only if $P^{\mu\nu}(\lie_{\xiv} g)_{\mu\nu}|_x=\xi_\mu H^\mu|_x=0$.

\subsection{Application to the Vaidya imploding spacetime}
The Vaidya spacetime (\ref{ds2}) has a proper Kerr-Schild vector field of type (\ref{KSVF}) relative to the null direction $\vec\ell$ given by $\xiv =\partial_v $ \cite{CHS}, because
$$
(\lie_{\xiv}g)_{\mu\nu}=2\frac{dm}{dv}\ell_\mu\ell_\nu \, , \hspace{1cm} (\lie_{\xiv}\ell )_\mu=0
$$
so that the function $h$ in (\ref{KSVF}) is $h=dm/dv\geq 0$. Note that $\xiv$ is hypersurface orthogonal, with the level function $\tau$ given by
\be
\xi_{\mu}dx^\mu =-Fd\tau = dr-\left(1-\frac{2m(v)}{r}\right) dv \label{tau}
\ee
Concerning the causal character of $\xiv$, notice that
$$
\xi_\mu\xi^\mu=-\left(1-\frac{2m(v)}{r}\right), \hspace{1cm} \ell_\mu\xi^\mu=-1
$$
so that $\xiv$ is future pointing on the region $\R=\r\cup\AH$, with $
\r:\hspace{2mm} r> 2m(v)$, timelike on $\r$ and null at the AH. 


Hence, the results in this section are applicable to $\xiv$: 
\begin{quotation}
{\em If the Vaidya spacetime (\ref{ds2}) satisfies (\ref{mdot}) and (\ref{mass}), then no closed generically f-trapped surface $S$ can be fully contained in the region $\r$. And the only ones contained in the region $\R:\,\, r\geq 2m(v)$ are the marginally f-trapped 2-spheres foliating the AH.}
\end{quotation}

Standard results \cite{HE,Wald} imply that no generically f-trapped closed surface can penetrate outside the EH. Given that EH is the past Cauchy horizon \cite{HE,S,Wald} of AH, EH=$H^-(\AH)$, the following conclusion follows: 
\begin{quotation}
{\em no closed generically f-trapped surface  can be fully contained in the region $D^-(\AH)$, so that they must penetrate the region $J^+(\AH)$ (given by $r\leq2m(v)$.)}
\end{quotation}
This agrees with theorem 4.1 in \cite{AG}. 

With regard to the boundary $\B$, we already know that closed trapped surfaces cross AH, and that the portion of EH within the flat region cannot be part of $\B$. But one can do even better and provide further restrictions on the location of $\B$. Put
$$
\tau_\Sigma\equiv \inf_{x\in AH} \tau |_x \, .
$$
It should be observed that $\tau_\Sigma$ is the least upper bound of $\tau$ on EH. The hypersurfaces $\tau=\tau_c$ are spacelike everywhere (and approaching $i^0$) if $\tau_c<\tau_{\Sigma}$, while they are partly spacelike and partly timelike, becoming null at AH, if $\tau_c>\tau_{\Sigma}$.
The location of the spherically symmetric hypersurface $\Sigma$ depends on whether $\frac{dm}{dv}(0)>1/8$ or not. In the former case, $\Sigma$ does enter into the flat region. It may not be so in the other cases. $\Sigma$ is shown in figure \ref{fig:fig1} for the particular case with (\ref{mass2}) and $\mu >1/8$. Notice that a characterization of $\Sigma$ is: the {\em last} hypersurface orthogonal to $\xiv$ which is non-timelike everywhere. 

$\Sigma$ is a relevant spacetime object because:
\begin{quotation}
{\em No closed generically f-trapped surface penetrates the region with $\tau<\tau_\Sigma$.}
\end{quotation}
The proof of this result uses the fact that the closed set above EH and below $\Sigma$ 
is contained in the region $\R$ where $\xiv$ is  future pointing so that any compact $S$ entering there will reach a minimum. But then one checks that this minimum should be local \cite{BD,BS}. The results of this section imply then that $S$ cannot be generically f-trapped.

The hypersurface $\Sigma$ is a past limit for f-trapped closed surfaces. In fact, they cannot even touch $\Sigma$, as follows from the fact that $\tau$ is decreasing along a connected path in $S$. Thus, finally we deduce that 
\begin{quotation}
{\em  all f-trapped closed surfaces lie in 
$\tau >\tau_\Sigma$ and have points with $r<2m(v)$.} 
\end{quotation}
Thus, $\B\subset \{\tau \geq\tau_\Sigma\}\cap \{r\geq 2m(v)\}$.

\section{The general spherically imploding spacetime}
The previous results can be generalized to the general imploding spherically symmetric spacetime with an asymptotically flat end.
The line-element can be written as
\be
ds^2=-e^{2\beta}\left(1-\frac{2m(v,r)}{r}\right)dv^2+2e^\beta dvdr+r^2d\Omega^2 \label{gds2}
\ee
where now $\beta(v,r)$ and the mass function $m(v,r)$ depend on $r$ and the null advanced time $v$. The spherically symmetric apparent 3-horizon AH is a marginally trapped tube defined by
$$
\AH : \hspace{1cm} r-2m(r,v)=0\, .
$$
AH is spacelike, null or timelike according to whether $
\left. \frac{\partial m}{\partial v}\left( 1-2\frac{\partial m}{\partial r}\right)\right|_{AH}$
is positive, zero or negative.
As before, let us define the region where the round 2-spheres are untrapped $
\r: \,\, r-2m(v,r)>0$. 

\begin{figure}[!ht]
\includegraphics[width=7.9cm]{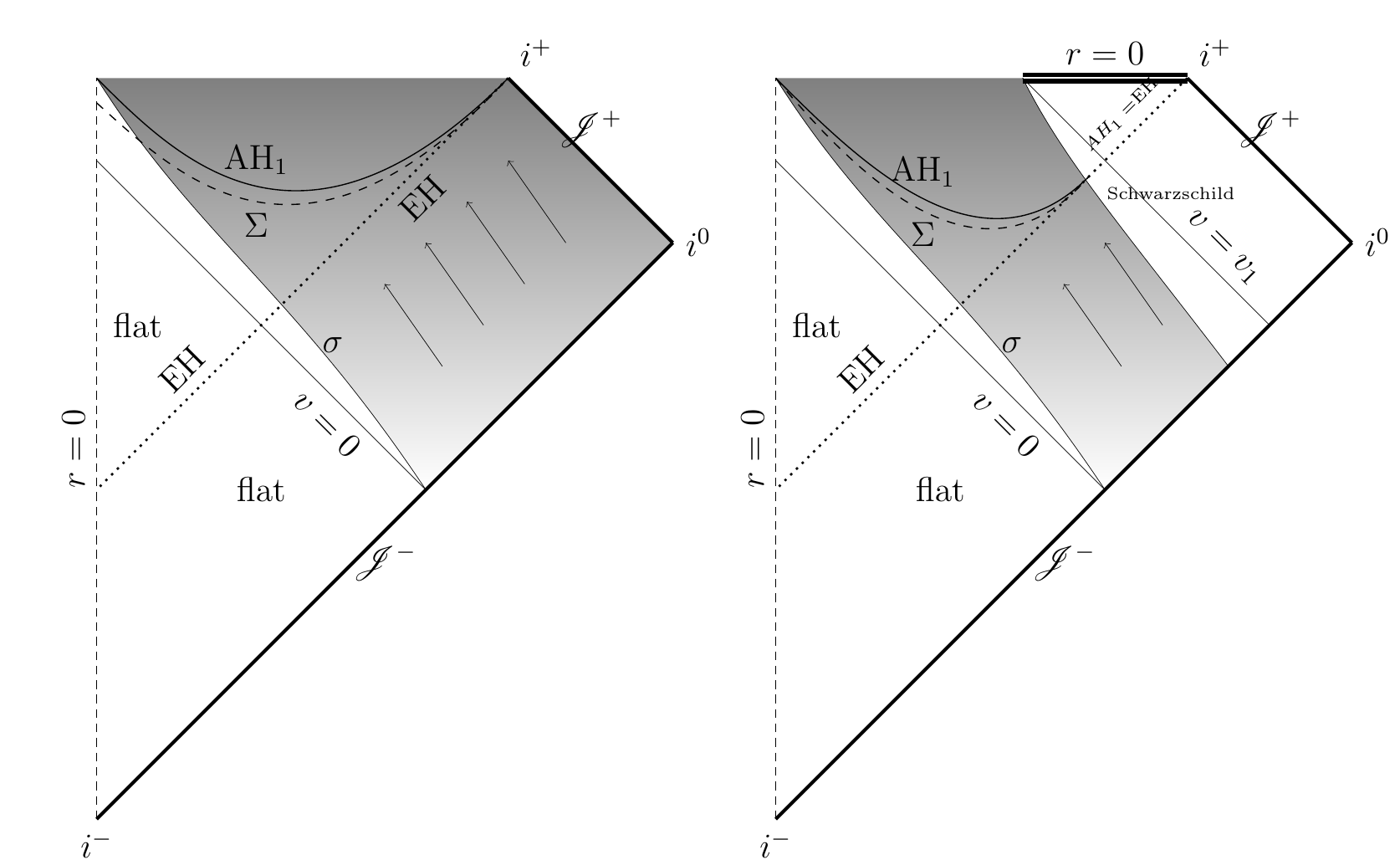}
\hspace{-0.5cm}
\includegraphics[width=7.9cm]{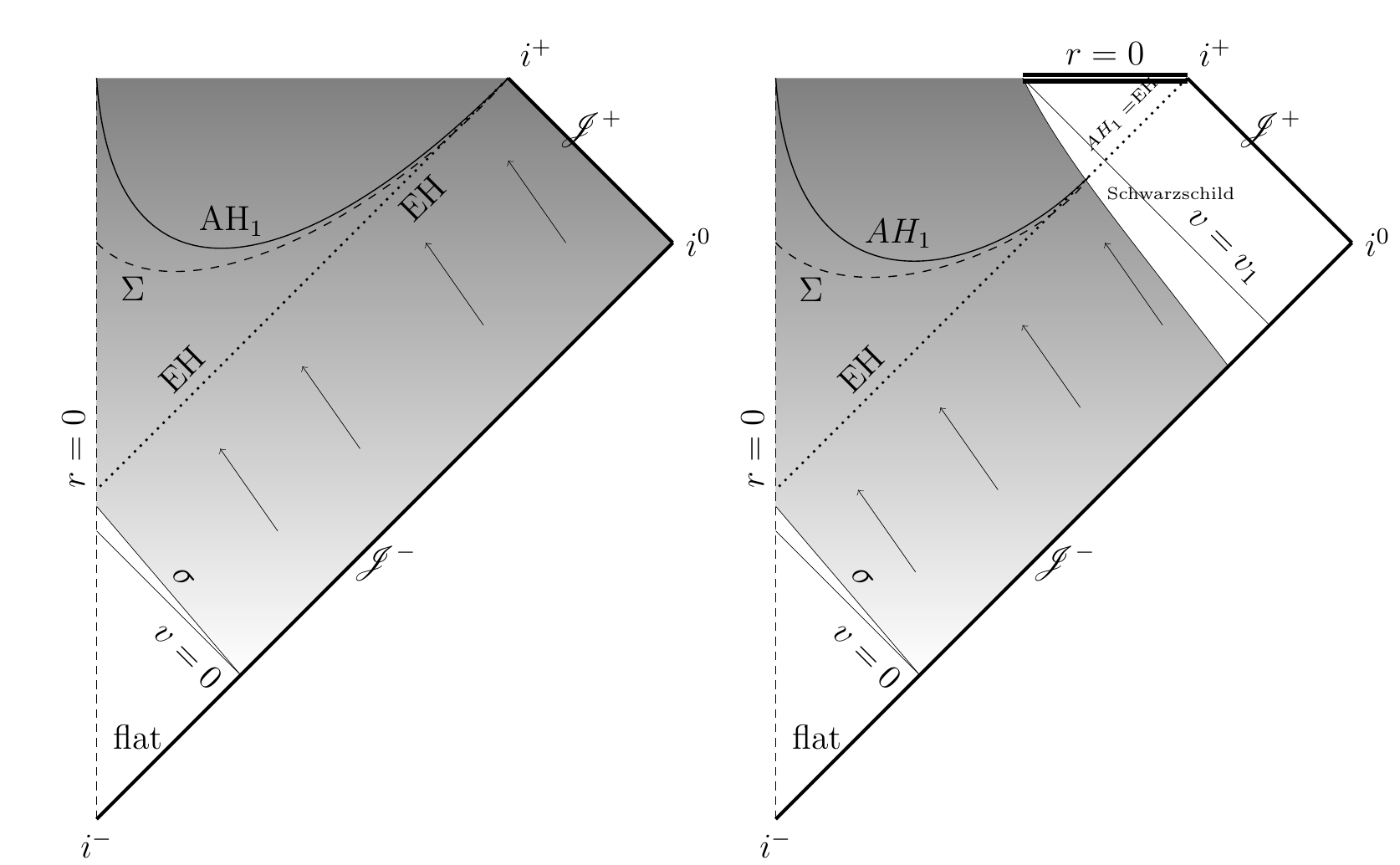}
\caption{Conformal diagrams of (\ref{gds2}) for the cases with $m(v,r)<M$ everywhere (first and third) or with $m(v,r)=M$ in some open asymptotic region (second and fourth). The spacetime is flat below $v=0$. EH may start developing in the flat region or not. The hypersurface $\sigma$ separating the flat portion and the rest of the spacetime cannot be spacelike. The shaded regions have non-vanishing energy-momentum. The connected component $\AH_1$ approaches EH either asymptotically or at some finite value of $v\leq v_1$ and $r=2M$. The first collapsing shell $\sigma$ may lead or not to the formation of a singularity depending on the properties of the mass function $m(v,r)$. The future evolution of the spacetime is thus left open for the shaded regions, and the third and fourth diagrams describe cases where there is a regular centre of symmetry in the non-flat non-empty region. $\AH_1$ is spacelike when approaching EH, but this is not necessarily so in other regions: in the last two diagrams, $\AH_{1}$ is timelike close to the left upper corner. The hypersurface $\Sigma$ puts a limit to the past on the possible location of f-trapped closed surfaces and, therefore, of $\B$. This hypersurface may enter the flat region (an example is the first diagram) or not. If the inflow of mass stops and there is a final Schwarzschild region of mass $M$, then $\Sigma$ merges with the EH and $\AH_{1}$ at $r=2M$ (second and fourth diagrams). The diagrams shown are only relevant possibilities, and many other possible combinations of the mentioned features are also feasible.}
\label{fig:fig2}
\end{figure}

Assume that the total mass function is finite and that there is an initial flat region:
$$
m(v,r)=0 \hspace{3mm} \forall v<0; \hspace{1cm} \forall v>0\,\left\{\begin{array}{c}
0\leq m(v,r)\leq M <\infty \\
m(v,r)\not\equiv 0
\end{array}\right.  .
$$
Then there is a regular $\scri^+$ and associated event horizon EH \cite{D}. $\AH_1$ denotes the connected component of $\AH$ associated to this EH. It separates the region $\R_1$, defined as the connected subset of $\r$ containing the flat portion, from a region containing f-trapped 2-spheres. The dominant energy condition is also assumed, and furthermore the matter-energy is incoming so that $\partial m/\partial v\geq 0$ on  $(\R_1\cup \AH_1)\cap J^+(EH)$. Under these assumptions, $\AH_1$ will eventually be spacelike (actually achronal) and asymptotic  to (probably merging) the EH \cite{Wil}. The relevant Penrose diagrams are shown in figure \ref{fig:fig2}.

\subsection{The hypersurface $\Sigma$}
The spacetime (\ref{gds2}) does not have a Kerr-Schild vector field in general, but we can insist on using $\xiv=\partial_v$, called the Kodama vector \cite{Ko}. This is hypersurface orthogonal with the level function $\tau$ defined by
\be
\xi_{\mu}dx^\mu =-Fd\tau = e^{\beta} dr-e^{2\beta} \left(1-\frac{2m(v,r)}{r}\right) dv \label{gtau} .
\ee
Its norm is 
$$
\xi_\mu\xi^\mu=-e^{2\beta} \left(1-\frac{2m(v,r)}{r}\right), \hspace{1cm} \ell_\mu\xi^\mu=-1
$$
where now $\ell_{\mu}dx^\mu=-e^\beta dv$, $\vec\ell =-e^{-\beta}\partial_{r}$, 
so that $\xiv$ is future-pointing timelike (null) on $\r$ (at $\AH$).

As in the Vaidya case, set $\tau_\Sigma\equiv \inf_{x\in AH_1} \tau |_x$ and define $
\Sigma\equiv \{\tau =\tau_\Sigma\}$.
The properties and characterization of $\Sigma$ are the same as in the Vaidya case.
Concerning its location, this depends on whether $8\dot m_0>(1-2m'_0)^2$ or not. 
Here $\dot m_0$ and $m'_0$ are the limits of $\frac{\partial m}{\partial v}$ and $\frac{\partial m}{\partial r}$ when approaching $(v=v_0^+,r=0)$, respectively, $v_{0}\geq 0$ being the value of $v|_{\sigma}$ at $r=0$ (see figure \ref{fig:fig2}).
In the former case, $\Sigma$ does penetrate the flat region, but it may not be so in the other cases  \cite{BS1}. Some possibilities have been shown in figure \ref{fig:fig2}.

The Lie derivative of the metric with respect to the Kodama vector can be easily computed 
$$
(\lie_{\xiv} g)_{\mu\nu} =e^{2\beta}\frac{2}{r}\frac{\partial m}{\partial v}\ell_\mu \ell_\nu -\frac{\partial\beta}{\partial v}\left(\ell_{\mu}\xi_{\nu}+\ell_\nu\xi_\mu \right)
$$
and this can be seen to be sufficient so that the fundamental results ---concerning the non-existence of a minimum of $\tau$ and related--- hold. Therefore, one can obtain the following important results:
\begin{itemize}
{\em 
\item No closed generically f-trapped surface can penetrate the region  $\tau<\tau_\Sigma$.
\item No closed f-trapped surface can enter the region $\tau\leq \tau_\Sigma$.
\item The minimum $\tau_m$ of $\tau$ on a closed f-trapped $S$ is always attained within $r\leq 2m(v)$,
\item furthermore, is $S$ happens to cross $\AH_1$, then $
\tau|_{S\cap \R_1} > \hat\tau_m > \tau_m$ and $ r|_{S\cap \R_1} < \hat r$
where $\hat\tau_m$ is the minimum value of $\tau|_S$ on $\AH_1$,
and $\hat r$ is the value of $r$ at the 2-sphere $\hat\varsigma \equiv \{\tau=\hat\tau_m\}\cap \AH_1$.}
\end{itemize}
As before, these properties of $\Sigma$ provide strong restrictions on the possible locations of the boundary $\B$. This is analyzed in more detail in the next section. 

\section{The boundary $\B$ in spherical symmetry}
Start by defining \cite{Hay,BS1} the future-trapped region $\mathscr{T}$
as the set of points $x\in \varietat$ such that $x$ lies on a closed f-trapped surface. $\mathscr{T}$ is an open set, as follows from the application of the formula for the variation of the null expansions (e.g. \cite{AMS} or many of references therein). However, 
$\mathscr{T}$ is not necessarily connected.

Denote then by $\B$ the boundary of the f-trapped region: $\B \equiv \partial \mathscr{T}$. This is related to the ``trapping boundaries'' in \cite{Hay}.
$\B$ being the boundary of an open set, it is a {\em closed} set without boundary. 
Moreover $\B \cap \mathscr{T}=\emptyset$. Observe that $\B$ divides the spacetime in two separate portions, because $\B$ is also the boundary of the untrapped region defined by the set of points $x\notin \mathscr{T}$. Again, $\B$ is not necessarily connected.
The connected component of $\B$ associated to $\AH_1$ will be denoted by $\B_1$.
It is important to remark that $\B$ is a genuine spacetime object, independent of any foliations or initial Cauchy data sets. Hence, $\B$ is basically different from the boundary of f-trapped surfaces contained in given slices and studied, e.g., in \cite{AM}. 

The previous properties are independent of spherical symmetry. If this symmetry is assumed, then one has: 
\begin{quotation}
{\em in arbitrary spherically symmetric spacetimes, $\mathscr{T}$ and $\B$ have spherical symmetry. Actually, $\B$ (if not empty) is a spherically symmetric hypersurface without boundary.}
\end{quotation}

Set $\tau_\B\equiv \inf_{x\in \B} \tau |_x$
where $\tau =$const. are the level hypersurfaces of the Kodama vector $\xiv$.
The following important results hold \cite{BS1}:
\begin{itemize}
{\em 
\item The connected component $\B_1$ does not have a positive minimum value of $r$,
\item moreover $\tau_\B = \inf_{x\in \B_1} \tau |_x=\tau_\Sigma$,
\item $\B_1\subset (\R_1\cup \AH_1) \cap \{\tau \geq\tau_{\Sigma}\}$,
\item $\B_{1}$ merges with, or approaches asymptotically, $\Sigma$, $\AH_1$ and EH in such a way that $(\B_1\backslash EH)\cap \AH_1=\emptyset$.
\item $\B_1\backslash$EH cannot be tangent to a $\tau=$const. hypersurface, so that 
$\tau$ is a monotonically decreasing function of $r$ on $\B_1\backslash$EH.
\item In particular, $\B_1\cap (\Sigma\backslash EH)=\emptyset$.
\item $\B_1$ cannot be non-spacelike everywhere. And it is spacelike close to the merging with $\Sigma$ and EH.}
\end{itemize}

\begin{figure}[!h]
\includegraphics[width=12cm]{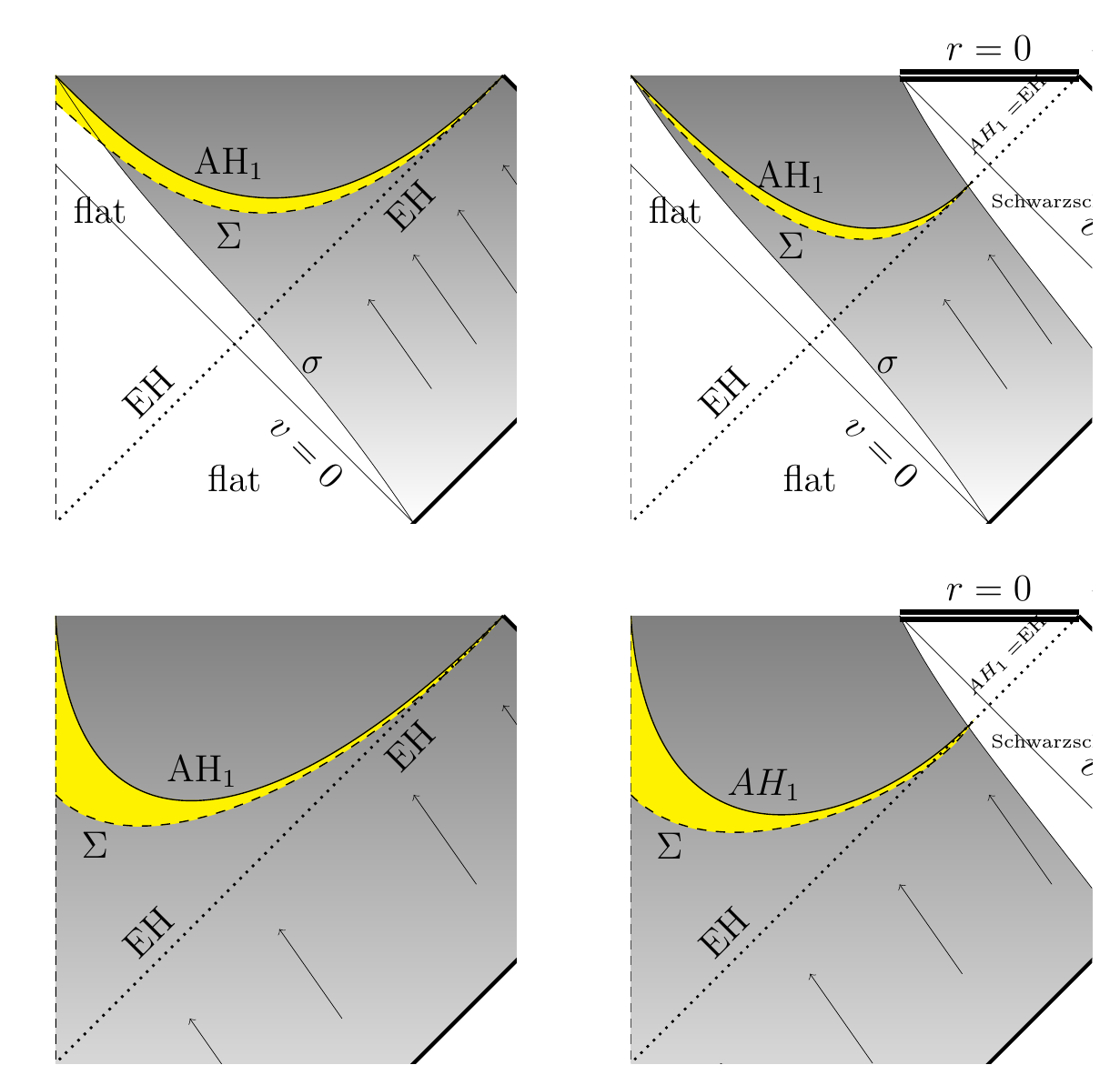}
\caption{These are enlargements of appropriate regions for the four cases shown in figure \ref{fig:fig2}. The boundary $\B_{1}$ must lie in the yellow region, and it cannot touch its upper and lower limit ---given by $\AH_1$ and $\Sigma$--- away from EH. If there is a Schwarzschild region with mass $M$ then $\B_{1}$ coincides with EH (and $\Sigma$ and $\AH_{1}$) there.}
\label{fig:fig3}
\end{figure}

These results prove that $\B_1\backslash EH$ must be placed strictly above $\Sigma$ and strictly below $\AH_{1}$. The allowed region for $\B_{1}$ is shown in figure \ref{fig:fig3} for several possibilities of interest.

Observe that $(\B_1\backslash EH)$ is entirely contained in the region $\R_{1}$, that is to say, in the region with $r>2m$ so that $\xiv$ is timelike and future-pointing there. Given that, from the fundamental results, no generically f-trapped closed surface can be contained in that region, it follows that 
\begin{quotation}
{\em $(\B_1\backslash EH)$ cannot be a marginally trapped tube, let alone a dynamical or future outer trapping horizon.}
\end{quotation}
Notice that the only closed marginally f-trapped surfaces that can be contained in $\B_1$ are those which are actually on its part $\B_1\cap$EH, if any. Observe also that this result implies that the notion of ``limit section'' in \cite{Hay} is generically non-existent, and thus theorem 7 in that reference is essentially empty in the sense that its assumptions are rarely met.

Let  
$$
\etav=\partial_v+e^{\beta}A\partial_r
$$
be the normal vector field to $\B_1$ and extend $A$ to be a function $A(v,r)$ on a neighborhood of $\B_1$, so that $\B_1$ belongs to a local foliation of hypersurfaces $\Sigma_t\equiv \{t=\mbox{const.}\}$, where $t$ is defined by ($G>0$)
\be
\eta_{\mu}dx^{\mu}=-Gdt =e^{\beta}dr-e^{2\beta}\left(1-\frac{2m}{r}-A\right) dv .\label{t}
\ee
By applying the reasonings used to derive the fundamental results to this hypersurface-orthogonal vector field on the regions $\B^s_1$ where it is spacelike ---for instance at the asymptotic region when $\B_1$ is about to merge with $\Sigma$ and $\AH_1$---, one can deduce that
\be
P^{\mu\nu}(\lie_{\etav} g)_{\mu\nu} |_{\B^s_1} \leq 0 \label{PK<0}
\ee
for {\em all} projectors of generically f-trapped surfaces tangent to $\B_{1}^s$ at some point. This puts severe restrictions on the boundary $\B_{1}$, see \cite{BS1}. In particular, 
\begin{quotation}
{\em $\B_1$ cannot have a positive semi-definite second fundamental form at any point where it is spacelike.}
\end{quotation}
Actually, the condition (\ref{PK<0}) is much more restrictive than this because it has to hold for {\em all} mentioned projectors. The combination of this with the spherical symmetry leads in turn to severe restrictions on the second fundamental form of $\B^s_1$ ---more generally, on the projection of $\lie_{\etav}g$ to $\B_1$, be this spacelike or not--- and its eigenvalues. This is work in progress \cite{BS1}.

\section{Conclusions and outlook}
The main conclusions are: 
\begin{itemize}
\item Closed trapped surfaces can penetrate flat portions of spacetime.
\item Closed trapped surfaces are highly non-local, a manifestation of the teleological character of the event horizon.
\item The boundary $\B$ seems to be a fundamental spacetime object, specially in asymptotically flat black-hole spacetimes. It defines the region where dynamical or future outer trapping horizons can exist.
\item $\B\backslash$EH does not include any portion of a marginally trapped tube. Actually, it does not contain any closed generically f-trapped surface.
\item The location of $\B$ has been severely restricted, and we are working on its intrinsic characterization.
\item Of course, one wishes to eventually give up spherical symmetry. In this sense
\begin{itemize}
\item The techniques used to define $\tau$ and to utilize it are completely general.
\item The main formula (\ref{main}), the general results on minima, etc. are also fully general.
\end{itemize}
\end{itemize}

\begin{theacknowledgments}
This paper is largely based on a collaboration with Ingemar Bengtsson \cite{BS,BS1}.
Comments from J.L. Jaramillo are acknowledged.
I thank the organizers of ERE-08 for the invitation, the Wenner-Gren Foundation for making this research possible, and the theoretical physics division at Fysikum in AlbaNova, Stockholms Universitet, for hospitality. Supported by grants FIS2004-01626 (MEC) and GIU06/37 (UPV/EHU).
\end{theacknowledgments}

\end{document}